\documentclass[epj,final]{svjour}

\usepackage{graphicx}
\usepackage{amsmath}
\usepackage{amsfonts}
\usepackage{amssymb}

\usepackage[numbers]{natbib}

\newcommand{\onlinecite}[1]{\hspace{-1 ex} \nocite{#1}\citenum{#1}}

\begin{document}

\bibliographystyle{unsrt}

\title{Resonant plasmon-phonon coupling and its role in magneto-thermoelectricity in bismuth.}
\author{P. Chudzinski}
\institute{DPMC-MaNEP, University of Geneva, 24 Quai
Ernest-Ansermet CH-1211 Geneva, Switzerland}

\institute{Institute for Theoretical Physics, Center for Extreme
Matter and Emergent Phenomena, Utrecht University, Leuvenlaan 4,
3584 CE Utrecht, The Netherlands}

\abstract{Using diagrammatic methods we derive an effective interaction
between a low energy collective movement of fermionic liquid
(acoustic plasmon) and acoustic phonon. We show that the coupling between
the plasmon and the lattice has a very non-trivial, resonant
structure. When real and imaginary parts of the acoustic plasmon's
velocity are of the same order as the phonon's velocity, the
resonance qualitatively changes the nature of phonon-drag. In the
following we study how magneto-thermoelectric properties are
affected. Our result suggests that the novel mechanism of energy
transfer between electron liquid and crystal lattice can be behind
the huge Nernst effect in bismuth.}

\date{\today}

\maketitle

\section{Introduction}

The thermoelectric signal, an electric response of a material upon
applying a temperature gradient, encodes how entropy is
transferred into the electronic system. The discovery of these
highly non-trivial transport coefficients dates back to
mid-nineteenth century, but recently this field enjoys renewed
interest of a broad scientific community. In fundamental research
it is a valuable tool to probe electron-hole
correlation\cite{Pourret-Nernst-Cooperpair} or formation of
lattice symmetry breaking orders\cite{Cyr-Nernst-stripes}. Its
role can be appreciated by a sole fact that the ultra-cold atom
systems were used to build quantum emulators of this
effect\cite{Brantut-Nernst-emulator}. In the field of applied
physics, there is an increasing demand for better handling with
heat\cite{Bell-applied, Hermans-viewpoint-appl}, a by-product in
any electric system. Devices that work as solar generators were
built\cite{Kraemer-solar-gen}, efforts to increase figure of merit
are made using nano-engineering\cite{Poudel-material-nano-engee},
effects involving spin degree of freedom were put
forward\cite{Bauer-spin-calor}, and this is to name only some
remarkable works from last few years. Still fundamental
theoretical understanding of an origin of finite thermoelectric
signal is frequently unsatisfying.

One case, where source of thermoelectricity is intensively
studied\cite{Oudovenko-thermoel-DMFT}, is the Hubbard model, where
strong on-site correlations can produce Kondo features in low
energy spectrum. An accepted scenario is that by lowering the
temperature one strongly affects the spectral function of
electrons and this in turn gives rise to a significant
thermoelectric signal. Although the best known, the Kondo-physics
is not the only situation, when collective phenomena strongly
affect the low energy spectral function of fermions.

In this work we wish to investigate an influence of non-local
correlations, collective plasmon excitations on the thermoelectric
transport coefficients, which provides a very different
perspective. In particular we focus on the problem how a low
energy collective mode (when exist) affects energy transfer to a
crystal lattice. In other words, we wish to understand better the
influence of electron-electron interactions on the phonon-drag
effect. For many decades the phonon-drag mechanism was
successfully used and allowed researches to explain
thermoelectricity in many materials, even despite the fact that
theoretical understanding of mechanism as such is usually limited
to single particle picture with an adiabatic approximation used to
describe the coupling.

Bismuth, a material chosen as a good example to apply our theory,
is a semimetal with an extremely anisotropic Fermi surface. Tiny
Fermi pockets originate only because of a lattice distortion from
cubic to rhombohedral. This suggest that coupling with lattice is
particularly strong\cite{Norin_Bi(p)DFT77, Tediosi_pressure}. The
extreme anisotropy of Fermi surface fulfills conditions for
appearance of a low energy collective excitation. Since the
acoustic plasmon dominates low energy dielectric function it has
to influence the low temperature phonon drag. The drag effects
usually manifest in thermoelectric effects and indeed with this
respect bismuth is quite intriguing. The Nernst signal in Bi is
one of the largest known in nature and, as explained in
Sec.\ref{ssec:Nernst-state-know}, it is beyond our
understanding\cite{Behnia-oscillating-N, Behnia-Science1st,
Behnia-Nat-com}.

The paper is organized as follows. In Sec.\ref{sec:model} we
introduce the band structure of bismuth and briefly present the
effective low energy theory. Next we give a diagrammatic
derivation of plasmon-phonon coupling in Sec.\ref{sec:plasm-phon}.
It is shown that the the low energy plasmons can couple with
acoustic phonons and the coupling vertex has a singular structure.
Next we show that the novel plasmon-phonon coupling mechanism does
manifest in magneto-thermoelectric phenomena and we focus on the
Nernst signal. Using results of Sec.\ref{sec:plasm-phon} we
estimate (in Sec.\ref{sec:Nernst-plasm}) a new phonon-drag
component. Finally in Sec.\ref{sec:disc} we discuss the
experimental relevance of our findings and indicate that several
discrepancies in the theoretical description of thermoelectricity
in bismuth can now be resolved.

\section{Model}\label{sec:model}

\subsection{Fermi liquid: results of band theory}

Bismuth is a semimetal whose Fermi surface consist out of three
electron pockets (at L points of Brillouin zone) and one hole
pocket (at T point). The pockets are tiny, $k_{F}$ are $10^3$
times smaller than the $T-L$ distance\cite{Allen_Bi-band95}. The
finite Fermi surface is due to a small (smaller than $3^\circ$)
lattice distortion which turns a cubic lattice into a rhombohedral
lattice, space group $A7$. This implies that the small density of
fermions available at lowest energies should be strongly coupled
with the lattice distortions.

The hamiltonian of those carriers close to the Fermi energy reads:
\begin{equation}\label{eq:ham_gen}
    H_{\rm fer}=H_{0}^{h}+\sum_{i=1}^3 H_{0}^{e_i}+H_{int}^{h-h}+H_{int}^{e-e}+H_{int}^{e-h}
\end{equation}
In the above $h$ denotes a hole pocket, while there are three
electron pockets indicated by $e_i$. The first two terms are the
free fermion kinetic energies, for instance: $H_{0}^{h}=\sum_k
E_h(\vec{k})(c^{(h)}_{\vec{k}})^\dag c^{(h)}_{\vec{k}}$, with $(c^{(\alpha)})^\dag_{\vec{k}}$ a creation operator of a fermion with momentum $\vec{k}$ in the $\alpha$ pocket. We approximate the kinetic energy
$E_{\alpha}(\vec{k})$ of each type of carrier by a a free fermion
dispersion relation. The peculiarity of bismuth are quite small
values of certain effective masses (of order $10^{-3}$) and large
anisotropy of the Fermi surface ellipsoids. An effective mass
ratio (in different directions) can be as large as $10^2$ and
between heavy holes and light electrons even $10^3$ (a detailed
values can be found in Ref.\citenum{Allen_Bi-band95}).
Extremely large ratios of masses of different carriers are present
for most orientations of a Bi crystal. When writing
Eq.\ref{eq:ham_gen} we focus on a parabolic part of fermionic
dispersion, excluding a non-parabolicity of the bands which
enters into the problem for energies above $20$meV. This simplification is
justified as we work in the lowest temperatures, in a low magnetic field (in Bi magnetic field can in principle cause inter-band transitions) and in the long-wavelength limit where our
interest is in a coupling with acoustic phonons (even high energy optical phonons in Bi have
energies $\approx 12$meV). Then the single particle excitations from the valence electron pockets (present at L points, at $\approx 40meV$ below $E_F$) are negligible. Nevertheless, one must keep in mind that the excluded high energy fermions contribute indirectly, for instance through a short-wavelength screening or a finite
lattice elasticity that determines phonon velocity or by contributing to the anomalous (orbital) diamagnetism of
bismuth\cite{Fuseya-Bi-rev'15}. In our framework, within renormalization group spirit, the effective low energy theory emerges as a result of downfolding (integrating out) higher energy degrees of freedom, these quantities are taken as a constant parameters. In a similar way, the non-parabolic dispersion may also strengthen the acoustic plasmon (see below in Sec.\ref{ssec:RPA}).

In hamiltonian Eq.\ref{eq:ham_gen} we have also introduced
hole-hole, electron-electron and electron-hole interactions.
Because the Fermi pockets are so tiny (dilute gas limit) the
Coulomb potential $V_{\rm Coul}(q)$ should be taken as a starting
point. As discussed in Ref.\onlinecite{moj-bismuth1} a good
approximation to capture the many-body effects in bismuth is the
Random Phase Approximation (RPA).

\subsection{Fermi liquid: results of RPA}\label{ssec:RPA}

The detail theory based on RPA applied to Bi was derived
elsewhere\cite{moj-bismuth1}. Here let us only recall that the
effective interaction between carriers $V_{\rm eff}(q)$ defined as
$V_{\rm eff}(q)= V_{\rm Coul}(q)/\epsilon(q,\omega)$ where
$\epsilon(q,\omega)$ is a dielectric function
$\epsilon(q,\omega)=1+(\Pi^{e}(q,\omega)+\Pi^{h}(q,\omega))$
obtained from polarizabilities $\Pi^{i}(q,\omega)$ of fermion
liquid components. The effective interaction has a resonant
structure with two plasmon poles, optical and an acoustic branch.
A well known optical plasmon has energy of order $10-20$ meV,
while for the other pole $\omega\rightarrow 0$, so it enters to
the lowest energy physics.

The fact that such collective excitation may exist in a two
component Fermi liquid is known already for
decades\cite{Pines_ac-plasm-classic62}. The key obstacle is that
the acoustic pole, fulfilling the condition $lim_{q\rightarrow
0}\epsilon(q,\omega=c_{\rm pl}q)=0$, can be weakly damped (not
overdamped) only for the material with a very anisotropic mass
tensor. This condition is perfectly fulfilled in bismuth. Moreover, even if we include the non-parabolic character of $E_e(\vec{k})$, this shall amount to include an extra term $\sim q$ in the denominator of $\Pi^{e}(q,\omega)$ thus effectively shifting the effective electron mass to even smaller values. This does not change qualitatively the picture and shall even make the acoustic solution more stable. The mode
has a zero sound character, but with an important difference, it
has a finite width for a finite momenta. The mode is present
because interaction between slower carriers are effectively
screened by the rapid ones, however its finite width implies that
screening is not complete (thus our model is distinct from the
Hubbard model). The relation between the damping and the mass
ratio can be understood if one realizes that acoustic plasmons can
survive the Landau damping as long as the linear plasmon
dispersion (determined by the velocity $c_{\rm pl}$, see
Eq.\ref{eq:ident}) is above the parabola of continuum of
electron-hole excitations of slower carriers. This condition gives
us an upper momentum threshold $k^*$ for the plasmons dominated regime (and, naturally, also energy $E^* = c_{pl}k^*$ and temperature $ E^*= k_B
T^*$ thresholds).

In this regime the hamiltonian Eq.\ref{eq:ham_gen} can be reduced
to an effective low energy hamiltonian $\widetilde{H}_{\rm fer}$
which consist of free electrons and holes coupled to the
collective bosonic excitation:
\begin{equation} \label{eq:Heff}
 \widetilde{H}_{\rm fer} = H_0^h + H_0^e + \sum_q \omega_q b^\dagger_q b_q +
  \frac1{\sqrt{\Omega}} \sum_{\alpha=e_i,h}\sum_q M_q^\alpha [b^\dagger_{-q} + b_q] \rho_\alpha(q)
\end{equation}
where $b^\dagger_q$ is a boson (plasmon) creation operator and:
\begin{equation} \label{eq:ident}
\begin{split}
 \omega_q &= c_{\rm pl} |q| \\
 M_q^{e0} &=  \left(\frac{3\pi^2 m_r c_{\rm pl}^3}{2 k_{Fr}^3}\right)^{1/2}
 |q|^{1/2}\\
 c_{\rm pl} &=\sqrt{V_{Fr}V_{Fs}/3}\\
\end{split}
\end{equation}
where $V_{Fr(s)}$ is the Fermi velocity of rapid (slow) carriers, $k_{Fr(s)}, m_{r(s)}$ are their respective Fermi wavelengths and effective masses, and $\rho_\alpha(q)=\sum_k c^{\dag}_{k+q}c_{k}$ is a density of
carriers from $\alpha$ pocket. The $M_q^{e0}$ is a $q$ dependent probability of a boson decay back into an e-h pair (this determines plasmon damping, or inverse lifetime, in the electronic system). In our problem both slow and rapid
carriers has a finite density at $E_F$ thus we must keep both in
Eq.\ref{eq:Heff}.

\subsection{Lattice: coupling with collective mode}\label{ssec:pl-phon-intro}

We proceed by incorporating lattice into our considerations. Then
the total hamiltonian reads:
\begin{equation}\label{eq:Ham_very_tot}
    H_{\rm tot}=\tilde{H}_{\rm fer} + H_{\rm ph} +H_{\rm coupl}
\end{equation}
In the above equation $\tilde{H}_{\rm fer}$ states for
Fermi-liquid hamiltonian as defined in (\ref{eq:Heff}). The
lattice hamiltonian $H_{\rm ph}$ is an acoustic phonon bath with a linear
spectrum:
\begin{equation}\label{eq:Ham_bath}
    H_{\rm bath}=\sum_{k} c_{ph}k a^\dagger_k a_{k}
\end{equation}

where $a^\dagger_k$ creates a phonon with momentum $k$.

$H_{\rm coupl}$ is, yet unknown, coupling between acoustic phonon and acoustic plasmon. A standard coupling, through
a displacement potential, a change of a single electron eigenenergies upon lattice distortion
(thus a direct coupling with the $H_0$ part of Eq.\ref{eq:ham_gen}), is
assumed to be known and not of interest in this study.

Another effect, that is very pronounced in
bismuth, is that a lattice distortion changes the volume of
electron/hole pockets (this is allowed in a semimetal). The effect is the strongest for the band bottom, the parabolic part of $E(k)$. When
$k\neq 0$, then modulation is not uniform in space, carriers have
to move to accommodate within the new energy landscape. Again, an extreme ratio of light/heavy carrier
masses, allows for a new phenomenon to occur. Different accelerations induce a
finite polarization of the two component Fermi liquid and this locally
produces a finite electric field $E_{\rm ind}$. Since the induced electric field potential $\nabla V_{\rm ind}= E_{ind}$
is proportional to lattice distortion, thus such coupling can be
interpreted as an effective piezo-electric interaction\cite{mahan_book}. It is known\cite{mahan_book} that $V_{\rm ind}$, which is phase shifted by $\pi/2$, does not interfere (at least up to a second order) with effects of displacement potential.

Physically, an additional contribution arises thanks to carriers' collective
motion -- numerous fermionic states buried deeply below $E_F$ are
able to couple with low energy lattice dynamics. Clearly, there is
no double counting: different groups of fermions drag phonons via
the standard mechanism or via the plasmonic channel. A possibility of direct coupling between phonons and plasmons
have been indicated already before, by a numerical evaluation of
integrals for $T=0$ charge RPA susceptibilities in low dimensional structures
\cite{Wendler_plasm-phon91, Wendler_plasm-phon86}. In the following section
we prove that a relevant scattering amplitudes are indeed non-zero
and evaluate their momentum dependence.

\section{Plasmon-phonon coupling}\label{sec:plasm-phon}

\begin{figure}
  \includegraphics[width=0.94\columnwidth]{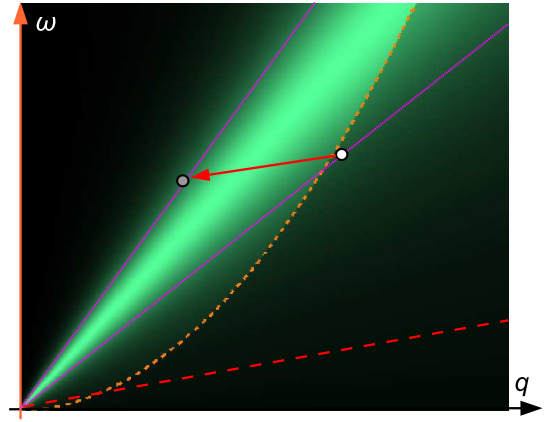}\\
  \caption{Phonon emission process (red vector) on energy-momentum plane. The green shading (bordered by two purple guide-lines) indicates weight of plasmon spectral function. Initial plasmon state (white dot) lies on the lower edge of the spectral function,
  while the final (grey dot) on the upper edge. Under the orange dashed line (a parabola) there is a zone where strong Landau damping by slow fermions sets in. Red dashed line shows phonon dispersion, it is parallel to the red vector. In the illustrated case one achieves $q_{ph}\approx q_{pl}$. }\label{fig:central-plot}
\end{figure}

Existence of the tiny Fermi pockets induced by distortion is one crucial property of bismuth. Below we derive the resonant coupling which reveals yet another prerequisite. As illustrated in figure Fig.\ref{fig:central-plot}, it is the existence of plasmonic and bosonic dispersions that fall sufficiently close on the momentum-energy plane. In a particular case of linear dispersion, it is a finite width of a plasmon that allows to compensate the difference of the two velocities. In bismuth, the desired acoustic plasmon exist, thanks to highly anisotropic tensors of effective masses for both electrons and holes.

\nopagebreak[4]
\onecolumn

\subsection{Bare plasmon-phonon vertex}\label{ssec:bare-coupling}

Bismuth crystal structure posses an inversion symmetry,
thus only the dynamically induced piezo-electric coupling is allowed. Then an issue is how a polarizability of Fermi liquid changes upon
emission/absorption of a phonon. When the
polarizability of the considered Fermi liquid has a singularity
for certain $\omega, k$ then it should dominate the response.  It
is then justified to use the plasmon-pole approximation.

\begin{figure}
  \centering
   \includegraphics[width=0.25\columnwidth, angle=-90]{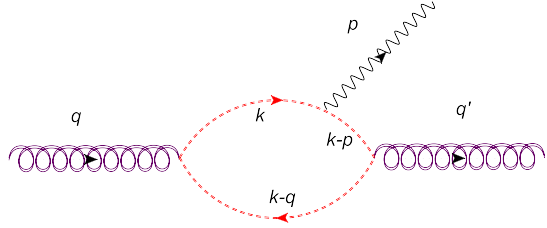}\\
   \caption{Emission of a phonon with momentum p (black zig-zag line) by a plasmon with momentum q (purple double solenoid line) with simultaneous emission of a plasmon with momentum q'. Red dashed lines are fermions' propagators
   inside an extracted bubble. This three leg bubble diagram represents the third order polarizability $\Pi^{III}(\omega,q,q')$.}\label{fig:bare}
\end{figure}

The simplest process is a direct transformation of a propagating
plasmon into phonon. The velocities of plasmons and phonons
differs by a factor four, thus energy-momentum conservation
prevents such events. 
In order to conserve energy and momenta the second plasmon must be
simultaneously emitted as it is shown in
Fig.\ref{fig:central-plot}. Then the processes to be considered
is plasmon destruction with simultaneous emission of another
plasmon and phonon which can be also seen as a coherent plasmon
backscattering on the lattice. This is shown on Fig.\ref{fig:bare} and described
by the following hamiltonian:

\begin{equation}\label{eq:tunnel}
    H_{p} = \sum_{q'}\daleth(q,q') b_q a^{\dag}_{q-q'} b^{\dag}_{q'} + {\rm h.c.}
\end{equation}

In a diagrammatic language, to evaluate $\daleth(q,q')$,
we extract a fermionic bubble from the acoustic plasmon
propagator\footnote{Since the bare
Coulomb line goes into an optical plasmon branch, an acoustic plasmon propagator
always has a fermionic bubble inside its propagator. One extracts a bubble
out of an infinite series from which the phonon
can be emitted (see Fig.\ref{fig:bare}).
Upon this operation the RPA series remains re-summable
and the plasmon propagator is still well defined, in fact it can
even increase its spectral weight thanks to translational symmetry
breaking by the lattice distortion. 
}. This bubble can emit a phonon
with the momentum $q_{ph}=q-q'$ and a plasmon with momentum $q_{pl}=q'$. The
emission of the plasmon, with properly tuned energy and momentum
is possible exclusively due to a finite life-time of
acoustic plasmon as illustrated on Fig.\ref{fig:central-plot}.

Then the $\daleth(q,q')$ is:
\begin{equation}\label{eq:plasm-phTunnel}
    \daleth(q,q')= M(q)V_{\rm ind}(q-q')M(q')\Pi^{III}(\omega,q,q')|_{\omega=0.8c_{pl}q}
\end{equation}
where the first term $M(q) \sim \sqrt{q}$ accounts for a plasmon
conversion into a bubble (particle-hole pair) and the
second term $M(q')$ is a conversion of a bubble back into plasmon, $V_{\rm ind}(q-q')$ is the induced piezo-electric potential which acts on a lattice, $\Pi^{III}(\omega,q,q')$ is a higher order polarizability of plasma. The
$\Pi^{III}(\omega,q,q')$ is a three leg bubble diagram: three
bosonic propagators leave a single bubble and one of these lines
is distinguishable from the others, see Fig.\ref{fig:bare}. We pick the $\omega$ value based on quantitative analysis given in
Fig.\ref{fig:central-plot}, but we have checked that the precise value does not affect the result, the resonant structure is always present. The three leg bubble diagram (drawn on
Fig.\ref{fig:bare}) can be expressed as follows:

\begin{equation}\label{eq:3-leg-def}
    \Pi^{III}(\omega_n,\vec{q},\vec{q'},\Omega_{2}^{ph})=\sum_{\omega_1,k}G(k-q,\omega_{1}-\omega_n)G(k,\omega_{1})
    G(k-p,\omega_{1}-\Omega_{2}^{ph})|_{p=q-q'}
\end{equation}

The finite temperature free fermion propagator is taken as
$G(k,\omega_{1})=(\omega_{1}-\epsilon_{k})^{-1}$ (with $\omega_1$
being fermionc Matsubara frequency) where $\epsilon_{\vec{k}}$ is
a single particle eigen-energy of the heavier carriers (it either
$E_h(\vec{k})$ or $E_e(\vec{k})$, in both cases an even function
of $\vec{k}$). In order to evaluate
$\Pi^{III}(\omega,q,q',\Omega_{2}^{ph})$ we proceed in a way very
similar to derivation of the (two leg) Lindhard function (see
App.\ref{app:3+4legs} for details). As a result at zero
temperature we get the following formula (see
App.\ref{app:3+4legs} for derivation):

\begin{multline}\label{eq:3-leg-integ}
\Pi^{III}(\omega,\vec{q},\vec{q'},\Omega_{2}^{ph}\rightarrow
0)=\int_{0}^{\vec{k}_{F}} d^3\vec{k}~\Bigg(\frac{1}{(\omega-\epsilon_{\vec{k}+\vec{q}}+\epsilon_{\vec{k}}+\imath\delta)(\omega-\epsilon_{\vec{k}}+\epsilon_{\vec{k}-(\vec{p}-\vec{q})}+\imath\delta)}+\\
\frac{1}{(\omega-\epsilon_{\vec{k}}+\epsilon_{\vec{k}-\vec{q}}+\imath\delta)(-\epsilon_{\vec{k}-\vec{p}}+\epsilon_{\vec{k}}+\imath\delta)}+
\frac{1}{(\omega+\epsilon_{\vec{k}+(\vec{p}-\vec{q})}-\epsilon_{\vec{k}}+\imath\delta)(\epsilon_{\vec{k}+\vec{p}}-\epsilon_{\vec{k}}+\imath\delta)}~\Bigg)|_{p=q-q'},\\
\end{multline}



\begin{figure}
  \centering
  \includegraphics[width=0.75\columnwidth]{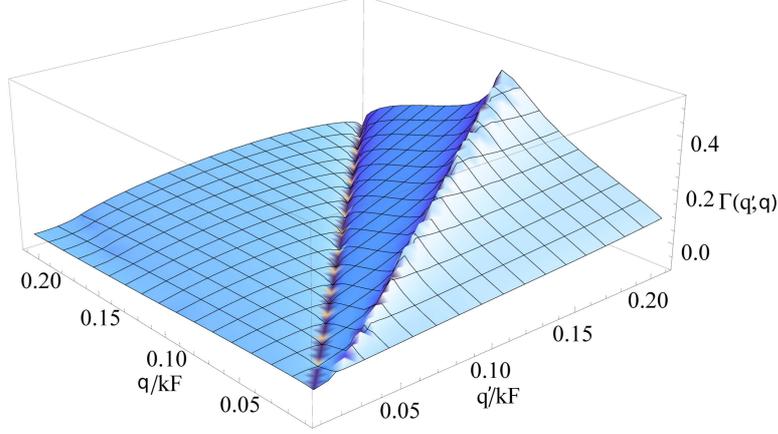}\\
  \caption{A real part of the scattering vertex $\daleth(q,q')$ on the q'-q plane: an incoming plasmon with momentum $q$ scatters into another plasmon with momentum $q_{pl}=q'$ and emits a phonon with momentum $q_{ph}=q-q'$.}\label{fig:vert1}
\end{figure}

The momentum integral can be, in some cases, performed analytically but formulas are rather long and thus
given in the appendix (App.\ref{app:3+4legs}). In the main text we only present the result, figure Fig.\ref{fig:vert1}, 
obtained for the specific set of parameters:
$Re[V_{pl}]=4Im[V_{pl}]=4V_{ph}$, which represents a particular case
of bismuth.


Figure Fig.\ref{fig:vert1} reveals a highly non trivial structure
of the scattering vertex $\daleth(q,q')$. There is a strong
resonance in momentum dependence spanned around the line $q'
\approx q/2$ which is equivalent to $q_{\rm ph}=q_{\rm pl}$. $Re[\daleth(q,q/2)]$ approaches values of
order $10^{-1}$, which on one hand implies that our perturbative
approach is reasonable, on the other that plasmon-phonon coupling
have a non-negligible strength.

Physically, existence of the resonance becomes clearer if one
considers an opposite process: when combined plasmon and phonon
produces a plasmon wave. This can be interpreted as a constructive
interference of two displacement waves. One then indeed expects a
resonance around $q_{\rm ph}=q_{\rm pl}$. Importantly, the
resonance condition restricts not only the value of wavevectors,
but also their direction, $\vec{q}_{ph}\parallel\vec{q}_{pl}$.

In Fig.\ref{fig:vert1} one easily notices that strength of
$\daleth(q,q/2)$ increases with $q$. The momentum
dependence of the resonant part is crucial for our further
reasoning in Sec.\ref{sec:Nernst-plasm}, so we investigate it in
detail. First we check, on log-log plots that
$\daleth(q,q/2)$ can be approximated quite well by a power
law, at least when we restrict ourselves to $q<0.1 k_F$ ($q^{*}<0.1
k_F$). For further calculations we take:

\begin{equation}\label{eq:power-depend}
    \daleth(q,q/2)= \bar{\daleth}q^{\alpha}
\end{equation}

The exponent varies (slightly) only
upon changing the ratio $c_{ph}/c_{pl}$: it is reduced from $0.57$ for
$c_{ph}/c_{pl}=1/4$ down to $0.52$ for $c_{ph}/c_{pl}=1/2$, while
changing other parameters even by factor five does not affect it.
At the same time the amplitude of $\daleth(q^*,q^*/2)$ may vary a few times.
Although the precise amplitude the coupling does depend on fine
details of the model, the exponent $\alpha\approx 0.5$ is universal enough
to allow for detection when fitting experimental results, which are routinely checked for the power-law behavior.

\nopagebreak
\twocolumn

\section{Nernst signal in bismuth}

A question is whether a plasmon-phonon resonance that we have
found in previous section is detectable. A good candidate, where
plasmon mediated phonon drag can manifest, are the thermoelectric
phenomena.

\subsection{Kelvin-Thomson relation}

We compute only a temperature dependence of the Nernst signal that is
characteristic for the novel mechanism. A straightforward '$S$'-approach, within the Nernst configuration,
is rather unpleasant from a theoretical viewpoint: one must
introduce variations of temperature within the sample, which is a
fundamental issue since the heat current will also contain
variations of chemical potential. Fortunately thanks to the
Onsager relation (known as Kelvin-Thomson relation in this case)
we can turn to a reciprocal coefficient, the Peltier coefficient
$\Pi_{xy}$. Then the Nernst signal $N\equiv S_{xy}$ can be
obtained from:

\begin{equation}\label{eq:K-O_relation}
    S_{xy}(T;B)=\Pi_{xy}(T;-B)/T
\end{equation}

from Eq.\ref{eq:K-O_relation} we see that Nernst signal can be
associated with entropy transport in the system. Now one compute
the Nernst coefficient using so called '$\Pi$'-approach, in the
Peltier configuration where a heat transfer is induced by an
electric voltage. Precisely, we will calculate a heat current
$\dot{Q_{y}}$ induced by a flow of electric current $j_{x}$ (under
a perpendicular magnetic field $B_{z}$). By definition the Peltier
coefficient $\Pi_{xy}$ is the proportionality constant:
$\dot{Q_{y}}=\Pi_{xy}j_{x}$ where $\dot{Q}$ is the time derivative
of heat diffused in y-direction. Thus, the price we need to pay in
the '$\Pi$'-approach is that we have currents flowing, a
non-equilibrium situation, in two perpendicular directions.

\subsection{Plasmonic contribution}\label{sec:Nernst-plasm}

First compute an amount of heat
transported by plasmons, or plasmonic clouds, accompanying
fermions. As suggested by experimental findings in
bismuth\cite{Tediosi_plasmaron07} we assume that in its movement
each electron is accompanied by a finite number of plasmons
$n_{pl}$, a so called plasmaronic picture. By analogy with
polarons\cite{Whitfield_polaron} we know that in average $\langle
n_{pl} \rangle= M_q$. We imposed a constant electric current
$j_x$, which can be associated with a voltage difference
$V_x=\rho_{xx}j_{x}$. In the presence of crossed fields $E_{x}$,
$B_{z}$ the electrons and holes move in the opposite directions
along the $x$-axis, but due to Lorentz force towards the same
electrode in $y-$direction. This is the $E\times B$ drift, with
constant velocity $u_y = E_x B_z$. It brings an extra entropy
(heat) to one side of conducting slab.

Then the heat transfer in y direction is:

\begin{equation}\label{eq:heat-y-plasm}
    \dot{Q_{y}}= j_x \rho_{xx} B_z \int dq^3 c_{pl}q M_q b(\beta c_{pl}q)
\end{equation}

with an integral which is very much like in a Debye model for
specific heat, except an extra $M_q$ term. As usual we extract
temperature dependence by rescaling the momentum $\tilde{q}=\beta
c_{pl}q $ (to get dimensionless argument of Bose-Einstein
distribution). Following these steps we find:

\begin{equation}\label{eq:Pelt-y-plasm}
    \Pi_{xy}(T;-B)\sim \rho_{xx} c_{pl} \bar{M} B_z T^{3.5}
\end{equation}

where $\bar{M}=M_q/\sqrt{q}$ is momentum independent. In passing
we note that $\Pi_{xy}\sim \rho_{xx}$ and this, for the case of Bi
where $\rho_{xx} \gg\rho_{xy}$, is in agreement with a general
tensorial relation between transport coefficients
$\hat{\Pi}=\hat{\alpha}\cdot\hat{\sigma}^{-1}$, where
$\hat{\alpha}$, $\hat{\sigma}$ are thermoelectric and electric
transport coefficients respectively. From Eq.\ref{eq:Pelt-y-plasm}
and Eq.\ref{eq:K-O_relation} it immediately follows that
$S_{xy}\sim T^{2.5}$.

In bismuth $c_{pl}$ is only slightly larger than $V_{Fs}$ (Fermi
velocity of slower carriers) and $\bar{M}$ is small. This implies
that an amplitude of plasmonic contribution to Nernst signal,
Eq.\ref{eq:Pelt-y-plasm} cannot be much larger than that of the
Fermi liquid. Then, in the lowest temperatures, a sole plasmonic
contribution ($\sim T^{2.5}$) cannot dominate the Fermi liquid
contribution ($\sim T^{1}$).

\subsection{Phononic contribution}\label{sec:Nernst-phon}

When the flow of fermion liquid induces lattice displacement the
main contribution to thermo-magnetic heat transfer will come from the large heat
capacity of the phonon gas. This is a realization of Gurievich
mechanism\cite{gurevich_classic61} in the presence of
collective excitations. We follow an approach frequently used for drag phenomena, see Ref.\cite{Matsuo_Nernst-phon09} for a specific case of bismuth,
and express $\Pi_{xy}$ as an extra heat that is transferred to the
lattice:
\begin{equation}\label{eq:Pi-drag-basic-formula}
    \dot{Q_{y}}=\int \frac{d^3q}{(2\pi)^3}c_{ph}
    q\cdot \hat{y} g(\vec{q})
\end{equation}
where $\hat{y}$ is a versor in y-direction and each phonon brings
(or takes) an energy $c_{ph}q$. The $g(\vec{q})$ is a displacement
of a phonon distribution $N_q$ from its equilibrium Bose
distribution $N_q^0$: $g(\vec{q})=N_q - N_q^0$ due to applied
fields. The $g(\vec{q})$ can be found from the Boltzman
equation for the steady state of phonons:

\begin{equation}\label{eq:Boltzman}
    \left(\frac{\partial N_q}{\partial t}\right)_{\rm force} + \left(\frac{\partial N_q}{\partial t}\right)_{\rm
    relax}=0
\end{equation}

where the two terms, that are balanced in the steady state, are
due to external forces e.g. due to $\vec{E},\vec{B}$ and relaxation e.g. due to
boundaries, impurities. Formally, Eq.\ref{eq:Boltzman} should be supplemented with three more equations, for plasmons and slow/fast fermions. Such system of partial differential equations could be treated only numerically. However, a few basic properties of bismuth, outlined in App.\ref{app:Boltz-aprox}, allowed us to restrict to Eq.\ref{eq:Boltzman} and obtain an analytic expression.

We take
$\frac{\partial N_q}{\partial t}_{\rm relax}= - g(q)/\tau_r$. This
is justified for the phonons whose distribution can relax
mostly by scattering on lattice imperfections (like
for instance twin boundaries known to be present in a rhombohedral
lattice).

The first term in Eq.\ref{eq:Boltzman}, due to the effective
fermion-phonon scattering, is estimated in Born approximation:

\begin{multline}\label{eq:Born-scattering}
  \left(\frac{\partial N_q}{\partial t}\right)_{\rm
  force}=\sum_{n,n'}[W^{em}(n',n)f_{n'}(1-f_{n})-\\
  W^{ab}(n',n)f_{n}(1-f_{n'})]
\end{multline}

where $f_{n'}$ is a shorthand notation for a Fermi-Dirac
distribution at energy $E_{n'}$. These energies does depend on
external forces, the fields $E_x$ and $B_z$, thus one can Taylor expand
Eq.\ref{eq:Born-scattering} up to $O(E_x)$. The procedure is
standard and we refer to original paper
Ref.\onlinecite{Fromhold-Boltz-formal-derived}. The Taylor expansion
brings a factor $\partial f(E)/\partial E$ which in the following (e.g. in
Eq.\ref{eq:total-heat-trans-sol}) limits integration to a 2D Fermi
surface (in all our considerations $k_B T \ll E_F$). The novel two
stage scattering enters through transition probabilities of phonon
absorption/emission
$$
W^{ab}(n',n)=b(\beta c_{ph}q)W(n',n);$$

$$W^{em}(n',n)=(b(\beta c_{ph}q)+1)W(n',n)
$$

where the bare probability $W(n',n)$ is evaluated from the
Kramers-Heisenberg formula:

\begin{equation}\label{eq:Kramers-Heisenberg}
    W(n',n)=\left(\frac{M_q
    \daleth(q,q/2)}{E(k+q)-E(k)-Im[c_{pl}]q}\right)^2
\end{equation}

where the difference between initial $n$ and final state $n'$ is a
phonon with momentum $q/2$. The intermediate state of the process, that is able
to emit a phonon, is a plasmaron. From
Eq.\ref{eq:Heff} a probability of creating an intermediate state,
a plasmon that accompanies fermion, is $M_q$. A probability that a
phonon will be
emitted, 
from Eq.\ref{eq:tunnel} is $\daleth(q,q/2)$. The last
formula, Eq.\ref{eq:Kramers-Heisenberg}, closes our system of
equations and by using Eq.\ref{eq:Born-scattering} together with
Eq.\ref{eq:Kramers-Heisenberg} we can solve Eq.\ref{eq:Boltzman}
for $g(q)$. This we can substitute back into
Eq.\ref{eq:Pi-drag-basic-formula} to find $\dot{Q_{y}}$, or after
dividing by $j_{x}=E_x/\rho_{xx}$ to find directly the Peltier
coefficient:

\begin{equation}\label{eq:total-heat-trans-sol}
    \Pi_{xy}(T)=\rho_{xx}\int_{0}^{q*} d^2 q c_{ph}q W(n,n')b(c_{ph}q/2)
\end{equation}

Comparing Eq.\ref{eq:Pi-drag-basic-formula} with
Eq.\ref{eq:total-heat-trans-sol} we see that

$$W(n,n')b(V_{ph}q/2)=g(\vec{q})$$

while from Eq.\ref{eq:Boltzman} we know that $g(q)\sim
\tau_p^{-1}$ where $\tau_p^{-1}$ is phonon scattering rate due to
interaction with plasmons. By comparison with results of
App.\ref{sec:self-ener} we see that the term $(M_q
\daleth(q,q/2))^2 b(c_{ph}q/2)$ in
Eq.\ref{eq:total-heat-trans-sol} is equal to the imaginary part of
the phonon self-energy (Eq.\ref{eq:scattering-rate-resul}) times
the plasmon spectral weight. The link between
Eq.\ref{eq:total-heat-trans-sol} and
Eq.\ref{eq:scattering-rate-resul} can be interpreted as a
generalized version of a Fermi golden rule with an inverse
scattering time $\tau_p^{-1}$ equal to $Im[\Sigma_{p}(q,\omega)]$.

We now use the fact that the denominator of
Eq.\ref{eq:Kramers-Heisenberg} scales linearly with momentum,
while momentum dependence of $M_{q}\sim\sqrt{q}$ and
$\daleth(q,q/2)\sim q^{\alpha}$ to find $W(n,n')\sim
q^{2\alpha-1}$. Being primarily interested in the temperature
dependence at the lowest energies we re-scale momentum
$\tilde{q}\rightarrow \beta q$ (the only temperature dependence in
Eq.\ref{eq:total-heat-trans-res} comes from an argument of the
Bose-Einstein distribution $\sim \beta q$). We find:

\begin{equation}\label{eq:total-heat-trans-res}
    \Pi_{xy}(T)= T^{2(1+\alpha)} \rho_{xx} A \int_{0}^{\tilde{q*}}d^2\tilde{q} c_{ph}\tilde{q} \tilde{q}^{2\alpha-1} b(c_{ph}\tilde{q}/2)
\end{equation}

where a momentum independent constant

$$A=\left(\frac{3\pi^2 m_r c_{\rm pl}^3}{2
k_{Fr}^3}\right)^{1/2}\bar{\daleth}/(c_{ph}-Im[c_{pl}])$$

The integral is known:

\begin{multline}\label{eq:Pi-integral-1}
    \int_{0}^{\tilde{q*}} d^2\tilde{q} c_{ph}\tilde{q}
    b(c_{ph}\tilde{q}/2)=\\ Li_2(\exp(c_{ph}\tilde{q^*}))+c_{ph}\tilde{q^*}\log(1-\exp(c_{ph}\tilde{q^*}))-(c_{ph}\tilde{q^*})^2/2
\end{multline}

where $Li_2$ is a di-logarithm function. In the limit of zero
temperature $\tilde{q^*}\rightarrow \infty$ and the right hand
side of Eq.\ref{eq:Pi-integral-1} becomes temperature independent.
Then, from Eq.\ref{eq:K-O_relation}, $N(T)=\Pi_{xy}(T)/T$ and we obtain the following temperature dependence for the Nernst
signal:

\begin{equation}\label{eq:N(T)-final}
    N(T)\sim T^{1+2\alpha}
\end{equation}

On the other hand for higher temperatures, when $k_B T \geq c_{ph}
q^*$, the temperature dependence of the right hand side of
Eq.\ref{eq:Pi-integral-1} matters. We approximate it by the last
term $\sim \tilde{q^*}^2 \equiv \sim T^{-2}$, to find that in high
temperatures $N(T)$ should slowly saturate to a constant value.
The regime of finite boson life-time is covered in our formalism,
the saturation profile will broaden and the saturation value
decrease when one includes finite imaginary velocity in the
argument of Eq.\ref{eq:Pi-integral-1}. What we have neglected so
far, is an intense Landau damping of plasmons that takes place
above the characteristic temperature $T^*$. This suppresses the
number plasmons available in the system. Since the plasmons are a
necessary ingredient of our drag mechanism, the $W(n,n')$ is also
suppressed and the $N(T>T^*)$ follows this decrease (see
Sec.\ref{sec:disc} for further discussion).

\section{Discussion, comparison with experiments}\label{sec:disc}

\begin{figure}
  \includegraphics[width=0.9\columnwidth]{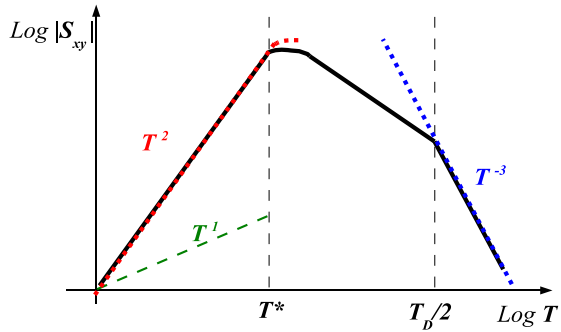}\\
  \caption{The predicted temperature dependence of the Nernst signal. The $T^2$ power law at the lowest temperatures is found from our novel plasmon-phonon coupling mechanism (red dotted line).
  The $T^1$ behavior (green dashed line) would be there for the Fermi liquid case. We also indicate the result of previous phonon-drag
  studies in in Ref.\cite{sugihara-old} (blue dotted line): in the highest temperatures a
  phonon drag with a power law decay $N\sim T^{-3}$ was predicted. For comparison an experimental result is sketched as a black solid curve. In experiments it was found that the lowest temperatures $T^2$ growth saturates around 3.5K ($\approx T^*$) while a regime where $N\sim T^{-3}$ is valid terminates around 50-80K that is approximately $\approx T_D/2$. }\label{S(T)}
\end{figure}

The predicted full temperature dependence of the Nernst signal is
shown on Fig.\ref{S(T)}. In
the lowest temperatures the consideration of the previous section
applies and the thermoelectric signal is $\sim T^{1+2\alpha}$.
Above the characteristic temperature $T^{*}$ (see
Sec.\ref{ssec:RPA}) the Coulomb interaction are no longer mediated
by collective excitations: fermions excited around $E_F$ are too energetic
to be followed by acoustic plasmons instead they cause strong
Landau damping.

Experimentally, in bismuth\cite{Behnia_Nernst07,
Behnia-oscillating-N}, and for the entire $Bi_{1-x}Sb_{x}$ family
of semimetals as well ($x<0.06$)\cite{Behnia_BiSb}, a power law
$N(T)\sim T^b$ is clearly observed, with $b\approx 2$, for
temperatures below $T \approx 3.5K$, a behavior that is clearly
different than the one expected for the Fermi liquid. The maximum
at 3.5K is strikingly similar to the upper limit of plasmonic-type
Baber resistivity\cite{Uher_Bi-rho77} and the previous
estimate\cite{moj-bismuth1} $T^{*}\approx 2\div 4K $. No other
characteristic energy scale of the material falls in this energy
range. Above $3.5K$ a decay of the Nernst signal was observed.

At higher temperatures, for $T>20K$ the Nernst effect was
explained in Ref.\cite{sugihara-old}. In that work a standard
single-fermion phonon drag was considered in the regime where the
phonon-phonon scattering dominates relaxation. The $N\sim T^{-3}$
decay was predicted, while the presence of temperature dependent
relaxation could be detected as deviation from this power law
$N\sim T^{-3+\delta}$ where $0<\delta<1$. This was successfully
compared with experimental data from Farag and
Taruma\cite{sugihara-old}. The invoked amplitude of electron-phonon coupling falls closely to
a recent \emph{ab-initio} estimate\cite{Hansen_def-pot94}. The $N\sim
T^{-3}$ was found for $T>80K$, while at lower temperatures $N\sim
T^{-2}$ fits better which is an indication that $1/T$ temperature
dependence of the relaxation rate is present and one needs to go
beyond standard electron-phonon coupling theory. This is what is
achieved in the present study.

We should note two recent theoretical studies of Nernst effect in bismuth, although these are primarily devoted to the high magnetic field regime. First work, Ref.\cite{Matsuo_Nernst-phon09}, by invoking only the phonon-drag mechanism, within the $"\Pi"$-approach, allowed to successfully explain the positions of peaks in the Nernst signal at high magnetic fields. Second work,
Ref.\cite{Sharlai-r(T)}, treated the same problem, but primarily focusing on the electronic, dissipationless contribution. It has also allowed to successfully explain the experimental signal, however a substantial phonon-drag contribution (of the same order as electronic part) needed to be accounted for. In both works the Nernst signal is proportional to longitudinal resistivity $\rho_{xx}$ and fermionic density, in agreement with the novel mechanism postulated here. In both works strong temperature dependence was found, for instance in Ref.\cite{Matsuo_Nernst-phon09} it is certainly stronger than linear dependence. The temperature dependence of phenomenological phonon-drag postulated (at low magnetic  fields) in Ref.\cite{Sharlai-r(T)} turns out to be one power faster than the dissipationless component, and falls close to $T^2$. This is in agreement with our Eq.\ref{eq:N(T)-final}, the main finding of our work. To complete this comparison, it should be clearly stated that in high magnetic fields the novel mechanism proposed here needs to be substantially modified: since then a magneto-plasmon has a quadratic, massive dispersion it will couple with an optical phonon, the amplitudon of the lattice distortion.

Furthermore, we compare the transverse (Nernst) and longitudinal (Seebeck)
signals. We use axes convention from Sec.\ref{sec:Nernst-plasm}.
In an ambipolar material, along the $x-$axis electrons and holes
move towards the opposite electrodes, thus accompanying them
plasmons are dragged in the opposite direction along this axis.
The longitudinal, Seebeck part of thermoelectric signal is then
expected to be small. Upon a presence of magnetic field, both
electrons and holes deviates in the y-direction towards the same
edge, the dragged plasmons and phonons can then build up a large
difference of temperatures. This implies that, at the lowest
temperatures, for plasmon-phonon drag $S_{xx}\ll S_{xy}$ and
indeed this is observed in the
experiment\cite{Behnia-oscillating-N}. The anomalous low
temperature signal is much stronger in the transverse (Nernst)
than in longitudinal (Seebeck) effect. In principle the
phonon-plasmon regime is limited also on the low temperature side,
by the Fermi liquid component $\sim T$. The amplitude of this
component is much smaller, and indeed such behavior was not
detected even down to mK temperature range in the transverse,
Nernst signal, while competition between different mechanisms has
been observed in the Seebeck signal\cite{Uher_Nernst_e-h78}.

Finally let us look at propagation of phonons. While they
experience mixing with plasmons, there are no point-like
collisions with fermions or defects. The drag process is driven by
the normal (not umklapp) scattering with other bosons which
implies that the only effect is the renormalization of phonon
velocity. In this low temperature regime, in the transport
experiment, the phonons can be detected as
ballistic\cite{Behnia_Nernst07}, but their velocity should be
slightly larger than expected. It is expected that heat transfer
is dominated by the bosonic bath and indeed ballistic transport
was found when measuring thermal conductivity of
bismuth\cite{Behnia_Nernst07}. Moreover, a sudden drop of thermal
conductivity was observed at $T\approx T^*$, which suggest a
profound link between the thermal transport and the
thermo-magnetic signal.

One may also consider more sophisticated inverse probes -- like
the change of sound propagation upon applying magnetic field.
According to the novel plasmon-phonon coupling proposed here they
can be affected. Indeed anomalously large oscillatory sound
attenuation (an analog of Shubnikov-de Hass) were experimentally
observed \cite{Kuramoto_attenu-e-h82, Kajimura_attenuation75,
yamada_acoust-magn65} again at temperatures of a few kelvins.
Detail study of these phenomena, as well as anomalous attenuation
of hot electrons in Bi, can provide further evidences of the novel
mechanism of coupling with a lattice.

\section{Conclusions}

In this work we have investigated influence of an acoustic plasmon
on a coupling between Fermi liquid and a lattice. We found quite
non-trivial momentum dependence of the coupling vertex $\daleth(q,q')$ with an
emergent resonance where plasmon-phonon coupling can be
significant. The resonance is reachable only thanks to a finite
imaginary part of the acoustic plasmon velocity. This is quite
distinct from the short range Hubbard-like models where the
zero-sound is infinitely sharp. Our model is dedicated to
extremely diluted regime, where non-local correlation effects are
dominant. One manifestation of the plasmon-phonon resonance is a
novel component of the phonon drag -- plasmon mediated phonon
drag. Bismuth is indicated as a likely material where such
coupling can take place and indeed we are able to provide an
interpretation for the huge low temperature Nernst signal in this
element.

There are other materials, for which our results are relevant. One prominent example are di-chalcogenides whose anisotropy of the
Fermi surface falls close to the one of bismuth. An unusual
charge density wave transition, that sometimes is ascribed to the
formation of an excitonic insulator\cite{Aebi_EI07}, takes place
in these materials. An issue of the role and nature of
electron-phonon coupling is a subject of intense ongoing
research\cite{Porer-exciton-lattice-separ}, electron-phonon vertexes with strong momentum dependence were
recently invoked\cite{Weber-Xray-qdep, Zenker-chiral-lattice} to
explain some of the peculiarities.

Our results may be useful from applied physics viewpoint, in low dimensional
(1D and 2D) nano-structures. In these structures plasmons are always acoustic ($\omega(q\rightarrow 0)\rightarrow 0$) and their
velocity (real and imaginary part) can be to some extend tuned. Our work shows that
by varying the plasmon's velocity one can significantly change the magnitude of
phonon drag. Two applications are possible, the first, mentioned
already in the introduction, is to optimize for a maximal
magneto-thermoelectric signal, in a desired direction, by enhancing the plasmon-phonon resonance. The second,
that stems from a growing interest in quasi-2D plasmonics, is to suppress the phonon-drag resonance in order to minimize looses due to non-radiative decay of acoustic plasmons.

\section*{Acknowledgments}

I would like to thank Thierry Giamarchi for many inspiring
discussions, especially in the initial stage of this project.

\section*{Author contribution statement}

The author solely performed all the calculations and wrote the manuscript.

\appendix

\section{Thermoelectricity in Bi, a state of
knowledge}\label{ssec:Nernst-state-know}

The Nernst signal measures a perpendicular electric voltage caused
by a temperature gradient (in the presence of magnetic field):

\begin{equation}\label{eq:Nernst-basic}
    N=\hat{\sigma}^{-1}\hat{\alpha}=\frac{\alpha_{xy}\sigma_{xx}-\alpha_{xx}\sigma_{xy}}{\sigma_{xx}^2+\sigma_{xy}^2}
\end{equation}

The Sondheimer cancellation between
Hall and transverse thermoelectric voltages, the numerator in
Eq.\ref{eq:Nernst-basic}, means that the signal is related to electron-electron interactions (the
non-interacting electrons propagator gives a zero
signal\cite{mahan_book}).  For a metal behaving according to a
semi-classical theory the Mott relation $\alpha \sim\partial
\sigma/\partial \omega$ is obeyed. The Mott relation express the
fact that non-zero thermoelectric power $\alpha_{ij}\sim \langle
J_i;J_j^E \rangle$ can appear only from the energy (momentum)
dependent part of self-energy\cite{Mahan-adiab-approx}. Despite these cancellations there are
attempts to ascribe the extremely large amplitude of the Nernst
signal solely to a standard Fermi liquid component. In
Ref.\onlinecite{Behnia_Nernst07} a very small value of the Fermi
energy and a very long electron scattering time were claimed to be
responsible and a following relation was given:
\begin{equation}\label{eq:N_in_Fermi}
    N=\frac{(\pi k_{B})^2
    T}{3}\frac{\partial\log(\sigma)}{\partial\omega}\approx \frac{\pi^2}{3}\frac{k_{B}^2 T}{m^\star}\frac{\tau}{E_{F}}
\end{equation}

The critical approximation done in Eq.\ref{eq:N_in_Fermi} are Mott
relation, assumption $\sigma_{xx}\gg \sigma_{xy}$ (and obviously
the same for $\hat{\alpha}$), Taylor expansion of a logarithm and
furthermore $\partial \sigma/
\partial \omega = \tau^{0}/E_F$, where a hidden assumption about
the relaxation time $\tau^{0}$ (which is supposed to approximate
the electrons scattering vertex) being a constant. The low energy
effective theory, Eq.\ref{eq:Heff} produces fermionic self-energy
that strongly depends on energy-momentum. This immediately implies
that Eq.\ref{eq:N_in_Fermi} does not hold. Even if one decides to
forget about this important component of self-energy, in a system
with an extremely small $E_F$ like Bi the band curvature matters
esp. when approximating the derivative (Eq.\ref{eq:N_in_Fermi}) by
the ratio, thus Eq.\ref{eq:N_in_Fermi} should be thought as an
upper limit for $N$. Strikingly, $N$ estimated by Eq.\ref{eq:N_in_Fermi}
is one order of magnitude smaller that the observed ones
(it is clear from Fig.~2 in \onlinecite{Behnia_Nernst07}).
Moreover the reasoning so far neglects one further cancellation:
bismuth is a semimetal so the mobilities of electrons and holes
partially cancel, which furthermore reduces the value estimated
from Eq.\ref{eq:N_in_Fermi}.

According to Eq.\ref{eq:N_in_Fermi}$, N(T)$ is expected to scale
linearly with temperature and this law should dominate as
$T\rightarrow 0$. This is in a disagreement with experimental
findings, a power law with larger exponent has been found, see
Sec.\ref{sec:disc} for details. There is an additional puzzling
element discussed in Ref.\cite{Matsuo_Nernst-phon09}.
Experimentally \cite{Behnia_BiSb, Behnia-oscillating-N} one has
the remarkable hierarchy $\sigma_{xx}\gg\sigma_{xy}$ (which is
typical for semimetals) and $\alpha_{xx}\ll\alpha_{xy}$. The two
relations definitely cannot fit together when the Mott relation
holds. To summarize: we do not claim that the dissipative
component of thermoelectric signal is zero, but that it cannot be
the dominating one at low temperatures.

The other component of thermoelectric power, not accounted in the
Fermi liquid theory, comes from the phonon drag. Thus the phonon-drag processes
are never accounted for in the approximation Eq.\ref{eq:N_in_Fermi}. The Mott
relation, giving the first equality in Eq.\ref{eq:N_in_Fermi},
holds and gives all contributions to $N$ only provided that
fermions are scattered on a static potential (of any
origin)\cite{Mahan-adiab-approx}. Phonon-drag
mechanism does not fulfill this condition.

\section{Approximation of Boltzman equation}\label{app:Boltz-aprox}

Our approximation is that fermions are strongly coupled with plasmons and relax only through plasmon-phonon scattering. This can be justified by:

\begin{itemize}

\item in bismuth there is $\sim 10^4$
atoms per one electron (tiny Fermi pockets), then lattice has much larger number
of degrees of freedom than electron liquid. According to second law of
thermodynamics entropy will flow from fermionic liquid to phonon
bath.
\item in low temperatures available electronic wavelengths are four orders of magnitude larger than lattice parameters, thus scattering with point defects is negligible\footnote{note particularly long
elastic scattering time for electrons in bismuth}, on the contrary the resonant plasmon-phonon coupling is rather strong so all the energy transfer must be done through this scattering channel
\item in the dilute gas limit electrons collisions are rare, moreover electron-electron are long range (favoring forward scattering) and below $T^*$ all these interactions are mediated by acoustic plasmons (providing partial screening)
\item the above described screening, taken together with a substantial imaginary part of plasmons' velocity, justifies the plasmaronic picture
\item only fermions are coupled with $\vec{E},\vec{B}$ fields and one then assumes their slightly shifted distribution, as it is routinely done in a dilute limit of weakly doped semiconductors, Ref.\onlinecite{Fromhold-Boltz-formal-derived}
\item in the lowest temperatures phonon-phonon collisions (in particular backward) are rare, the strongest "force" phonons receive is from the fermion drag
\end{itemize}

\section{Self-energy}\label{sec:self-ener}

We compute
$\Sigma_{p}(q,\omega)$, the self-energy of plasmons moving through
the bath of phonons. Upon its propagation, a plasmon with
momentum-energy $(q,\omega)$ emits and then reabsorbs a phonon
with momentum-energy $q',\omega_{ph}$. $\Sigma_{p} $ reads:

\begin{multline}\label{eq:self-Ep}
    \Sigma_{p}(q,\omega)=\int d\vec{q'}\sum_{\omega_1}\daleth(\vec{q},\vec{q'})\\
    G_{ph}(p=q-q',\omega_1)G_{pl}(q',\omega-\omega_1)\daleth_0(\vec{q},\vec{q'})
\end{multline}

where integral is over all possible momenta $q'$ of an emitted
phonon. The plasmon Greens function in spectral
representation:

$$G_{pl}(q,\Omega_{pl})=\bar{A}_{pl}(q)\left(\frac{1}{\Omega_{pl}-c_{pl}q}-\frac{1}{\Omega_{pl}+c_{pl}q}\right)$$

where $\bar{A}_{\rm pl}(q) = \left(\frac{3\pi^2 m_r}{k_{Fr}^3}\right) \frac{c_{pl}^3 |q|}2$. We perform summation over the bosonic Matsubara frequency
$\omega_1$ to get (after analytic continuation):

\begin{equation}\label{eq:self-ener-after-w1}
    \Sigma_{p}(q,\omega)=\int d\vec{q'}\frac{\daleth_0 \daleth \bar{A}_{\rm pl}(q)(b(\beta c_{ph}q')+b(\beta c_{pl}(q-q')))}{(\omega+c_{ph} q' \pm c_{pl} (q-q') +
\imath\delta)}
\end{equation}

Results of Sec.\ref{sec:plasm-phon} allow for
crucial simplifications. The coupling is resonant, which means
that for each incoming plasmon with momentum $\vec{q}_{pl}$ only a
phonon with a momentum $|\vec{q}_{ph}|=1/2|\vec{q}_{pl}^{i}|$ and
$\vec{q}_{ph}\parallel \vec{q}_{pl}^{i}$ fits the resonance
condition. This introduces $\delta(\vec{q'}-\vec{q}/2)$ and the
integral over $q'$ becomes trivial. On the resonance $2q'=q$ and this allows to extract
$1/q$ factor in the denominator of $\Sigma_{p}(q,\omega)$. This
leads to:

\begin{multline}\label{eq:self-ener-pl-ph-imagin}
    Im[\Sigma_{p}(q,\omega)]=\frac{2\daleth_0 \daleth \bar{A}_{\rm pl}(q)}{q}(b(\beta c_{ph,pl}q'))\\
    \delta((\omega/q-[c_{pl}-(c_{pl}\pm c_{ph})/2])^2+Im[c_{pl}]).
\end{multline}

We obtain an expected result (see Fig.\ref{fig:central-plot}) that
$Im[\Sigma_{p}(q,\omega)]\neq 0$ only for special values of
velocities $\omega/q = c_{pl}-(c_{pl}\pm c_{ph})/2$ and when there
is a finite spectral weight along these lines. The spectral
function $Im[G_{pl}(q,\Omega_{pl})]$ is a lorentzian centered
around $\omega/q = Re[c_{pl}]$ with width $\sim Im[c_{pl}]$. Then
$Im[\Sigma_{p}(q,\omega)]\neq 0$ only when $(c_{ph}) \leq
Im[c_{pl}]$. This is the physics presented in
Fig.\ref{fig:central-plot}, the resonance can be reached only if
plasmon spectral weight is broad enough and/or velocity of phonon
is close enough to velocity of plasmon.

About the momentum
dependence of $Im[\Sigma_{p}(q,\omega)]$: the $1/q$ factor cancels
with momentum dependence of $A_{\rm pl}(q,\omega
>0)\sim q$, so we are left with vertexes, the bare coupling has
dependence $\daleth_0 \sim q_{ph}^{\alpha}$, see
Eq.\ref{eq:power-depend}. Then, in the lowest order, the plasmon-phonon scattering rate has the following momentum dependence:
\begin{equation}\label{eq:scattering-rate-resul}
    \tau_p^{-1}=Im[\Sigma_{p}(q,\omega)]\sim q^{2\alpha} b(\beta c_{ph}q/2).
\end{equation}


\onecolumn

\section{Three leg bubble $\Pi^{III}(\omega,\vec{q},\vec{q'})$: details of calculations}\label{app:3+4legs}

In order to evaluate right hand side of Eq.\ref{eq:3-leg-def} we
begin with the sum over the internal Matsubara frequency of
fermions. Summing over $\omega_{1}$ in Eq.\ref{eq:3-leg-def} leads
to three terms corresponding to the three poles:
\begin{multline}\label{eq:3-leg-app1}
    \Pi^{III}(\omega,\vec{q},\vec{q'},\Omega_{2}^{ph})=
    \int_{0}^{\infty} d^3\vec{k}~\Bigg[\frac{f(\epsilon_{\vec{k}-\vec{q}})}{(\imath\omega_n+\epsilon_{\vec{k}-\vec{q}}-\epsilon_{\vec{k}})(\imath(\omega_n+\Omega_{2}^{ph})-\epsilon_{\vec{k}-\vec{q}}+\epsilon_{\vec{k}-\vec{p}})}+\\
    \frac{f(\epsilon_{\vec{k}})}{(\imath\omega_n-\epsilon_{\vec{k}}+\epsilon_{\vec{k}-\vec{q}})(\imath\Omega_{2}^{ph}-\epsilon_{\vec{k}-\vec{p}}+\epsilon_{\vec{k}})}+
    \frac{f(\epsilon_{\vec{k}-\vec{p}})}{(\imath(\omega_n-\Omega_{2}^{ph})+\epsilon_{\vec{k}-\vec{q}}-\epsilon_{\vec{k}-\vec{p}})(-\imath\Omega_{2}^{ph}+\epsilon_{\vec{k}}-\epsilon_{\vec{k}-\vec{p}})}~\Bigg],\\
\end{multline}
Each of these terms has a polynomial of the second order in the
denominator. There are no trivial cancellations between these
terms, in the following each of them will be treated separately. Since, $V_F\vec{k}_{F}\gg k_B T^*$ we are allowed to take zero temperature limit.
We make a proper shift of momentum variables to always have
$f(\epsilon_{k})$ in the numerators. Then the Fermi distributions
present in the numerators (Eq.\ref{eq:3-leg-app1}) simply restricts the
limits of momentum integrals to $\vec{k}=\pm \vec{k}_F$ at $T=0$,
we shall also use the fact that free-fermion dispersion is an even
function of $k$. Then we arrive at Eq.\ref{eq:3-leg-integ} in the main text.

Next we take the phonon frequency to zero ($\Omega^{ph}\rightarrow
0$). This significantly simplifies the calculation and is not
unreasonable as the phonon frequency is always the smallest energy
scale in the problem. Nevertheless we also performed auxiliary
calculations to check that for quantities of our interest, the
first order corrections $\sim\Omega^{ph}$ are indeed always one
order of magnitude smaller. From qualitative analysis (see
Fig.\ref{fig:central-plot}) we see that although finite
$\Omega^{ph}$ broadens the resonance, it also increases the
density of states of the plasmons involved in the process. The two
effects likely compensate.

A closer insight into Eq.\ref{eq:3-leg-def} provides mathematical
explanation of the resonance. If one make a shift by $q/2$ then
he/she can notice that the second and third fermionic poles gives
rise to terms like $(\epsilon_{k-p}-\epsilon_{k-q/2}+\Delta)$
(where small $\Delta$ is related to phonons dynamics) in the
denominators. In fact the one-loop master integrals, closely
related to the three leg bubble
$\Pi^{III}(\omega,\vec{q},\vec{q'})$, is known to have a resonance
feature.

In Eq.\ref{eq:3-leg-integ} one needs to perform a momentum
integral. Like for the case of Lindhard function
($\Pi^{II}(\omega,\vec{q})$ in our notation) the momentum integral
is easiest to perform in the cylindrical coordinates: if we take
$z\parallel \vec{q}$ then $\int d^3\vec{k}\rightarrow \int dk k
\sin\vartheta$ and $(\epsilon_{k+q}-\epsilon_{k})\rightarrow
(q^2/2m-qk\cos\vartheta)$, where $\vartheta\angle
(\vec{q},\vec{k})$. Firstly, one integrates over angle $\vartheta$
and then over $|k|$. The analytical formulas are available only
for the specific case when $p\parallel q'$, then
$\Pi^{III}(\omega,\vec{q},\vec{q'})\rightarrow
\Pi^{III}(\omega,q,q')$. However we did check by numerical
integration that the $\Pi^{III}(\omega,\vec{q},\vec{q'})$ indeed
has a broad maximum at around $\vec{p}\parallel \vec{q'}$ and is
close to zero for the angles $\angle (\vec{p},\vec{q}) \in
(\pi/3;2\pi/3)$. The following results is found for the three
terms of Eq.\ref{eq:3-leg-integ}:
\begin{multline}\label{eq:3leg-pol1}
  \Pi_1(q,p,w)=\frac{q (p-q)}{3 (p-2 q) (q (p-q)+w)}\Bigg(-\frac{\left((p-q)^2-w\right)^3 \log \left(-p^2+2 p q-p-q^2+q+w\right)}{(p-q)^3}-\log ((q-1) q-w)+2 q-\\
  \frac{\left((p-q)^2-w\right)^3 \log \left(-2 p q+(p-1) p+q^2+q-w\right)}{(p-q)^3}+\frac{\left((p-q)^2-w\right)^3 \log \left(w-(p-q)^2\right)}{(p-q)^3}+\log \left(q^2+q-w\right)+\\
  \log \left(-2 p q+(p-1) p+q^2+q-w\right)-\frac{w (p-2 q)}{q (p-q)}+\frac{\left((p-q)^2-w\right)^3 \log \left((p-q)^2-w\right)}{(p-q)^3}-\log ((p-q) (p-q+1)-w)-p+\\
  \frac{\left(q^2-w\right)^3 \log \left(q^2+q-w\right)}{q^3}+\frac{\left(q^2-w\right)^3 \log \left(-q^2+q+w\right)}{q^3}-\frac{\left(q^2-w\right)^3 \log \left(q^2-w\right)}{q^3}-\frac{\left(q^2-w\right)^3 \log \left(w-q^2\right)}{q^3}\Bigg)
\end{multline}
\begin{multline}\label{eq:3leg-pol2}
   \Pi_2(q,p,w)=-\frac{q}{3 (q (p+q)-w)}\bigg(-2 p^3 \tanh ^{-1}(2 p+1)-2 p^3 \tanh ^{-1}(1-2 p)-p-\log (-p-1)+\log (1-p)-\\
   \log \left(q^2+q-w\right)+\frac{\left(q^2-w\right)^3 \log \left(-\left(q^2-w\right)^2\right)}{q^3}-\frac{\left(q^2-w\right)^3 \log \left(-q^4+q^2 (2 w+1)-w^2\right)}{q^3}+\frac{w}{q}+\log ((q-1) q-w)-q\bigg)
\end{multline}
\begin{multline}\label{eq:3leg-pol3}
   \Pi_3(q,p,w)=-\frac{(p-q)}{3 \left(-p q+q^2+w\right)}\Bigg(\log \left(p^2-2 p q+p+(q-1) q+w\right)-\\
   \frac{1}{(p-q)^3}\bigg[\left((p-q)^2+w\right)^3 \Big(-\log \left(-p^2+2 p q+p-q^2-q-w\right)-\log \left(p^2-2 p q+p+(q-1) q+w\right)+\\
   \log \left(-(p-q)^2-w\right)+\log \left((p-q)^2+w\right)\Big)+2 p^3 (p-q)^3 \tanh ^{-1}(2 p+1)+2 p^3 (p-q)^3 \tanh ^{-1}(1-2 p)\bigg]\\
   -\log \Big(-2 p q+(p-1) p+q^2+q+w\Big)+\frac{w}{p-q}+\log (p-1)-\log (p+1)-q\Bigg)
\end{multline}
These results are actually close analogues of the so called \emph{"master
integrals"}, known from the loop expansion in the high energy
physics. Like the \emph{master integrals}, our result is also rather
complicated combination of logarithms and di-logarithm functions.
The di-logarithm functions are present in the more complicated cases: for the four
leg bubble $\Pi^{IV}(\omega,\vec{q},\vec{q'})$, for non-adiabatic
case (or propagators with finite life-times) or when $\vec{p}$ is
not parallel to $\vec{q}$.


\twocolumn

\bibliography{plasm}

\end{document}